 \newcommand{\RE}{(RE)\-Ba$_{2}$\-Cu$_{3}$\-O$_{7-\delta}$}
  \newcommand{\bitwo}{Bi$_{2}$\-Sr$_{2}$\-Ca\-Cu$_{2}$\-O$_8$}
 
              \newcommand{\bc}{\begin{center}}       
              \newcommand{\ec}{\end{center}}         
              \newcommand{\beq}{\begin{equation}}    
              \newcommand{\eeq}{\end{equation}}      
              \newcommand{\bitem}{\begin{itemize}}   
              \newcommand{\eitem}{\end{itemize}}     
              \newcommand{\bnum}{\begin{enumerate}}  
              \newcommand{\enum}{\end{enumerate}}    
              \newcommand{\noi}{\noindent}           
\documentstyle[prb,aps]{revtex}            
 \tighten
\input{psfig.tex} 
\begin{document}
 \setcounter{equation}{0}  \draft                 
\twocolumn[\hsize\textwidth\columnwidth\hsize    
          \csname @twocolumnfalse\endcsname     

\title{Irreversible shape distortions by flux-pinning-induced magnetostriction in hard type-II superconductors}
\author{T.H. Johansen, J. Lothe and  H. Bratsberg }
\address{Department of Physics, University of Oslo,
    P.O. Box 1048, Blindern, 0316 Oslo 3, Norway.}
\date{\today}
\maketitle
\widetext
\begin{abstract}

Exact analytical results are obtained for the flux-pinning-induced magnetostriction in
long cylinders of type-II superconductors placed in a parallel magnetic field.
New effects are found regarding the irreversible deformation  of the cross-sectional area: 
Whereas the magnetostriction of a circular cylinder is shape conserving, it is shown
that a square cross-section deforms with considerable distortion.
During a field cycle both concave, convex, and even more complicated distortions are predicted.
The strong implications for the interpretation of dilatometric observations  are pointed out.
The main results obtained for both the circular and square geometry 
are valid for any critical-state model,  $j_c = j_c(B)$, and we use 
the Bean model for illustration and discussion.

\end{abstract}
\vskip2pc]
\narrowtext
%
%
It was discovered by Ikuda and coworkers \cite{Ikuta-1}
 that high temperature superconductors (HTSCs),
when placed in a magnetic field, can respond by a giant magnetostriction.
Relative changes in sample size as large as $10^{-4}$ were observed in single crystals of \bitwo.
To explain this deformation, which is  about two  orders of magnitude larger than values reported for
conventional superconductors,\cite{Brendli} they considered 
the stress exerted on the crystal by pinned vortices.  Treating the crystal as an infinite slab
of half-width, $w$, placed in a parallel applied field, $B_a/\mu_0$,
the pinning-induced magnetostriction in the transverse direction was derived from
 the critical-state model to be 
\begin{equation}
   \Delta w/w =
  (2 c  \mu_0 w)^{-1} \;  \int_0^w \; [ B^2(x)-B^2_a ] \; dx  ,
\label{eq:1}
\end{equation}
where  $c$ is the elastic constant and $B(x)$ the local induction.
As evident from eq.(\ref{eq:1}), and also pointed out immediately by Ikuta et al.,
the  pinning-induced magnetostriction is intimately related to the  magnetization,
and consequently should display  similar  irreversible characteristics. 
Following the discovery, the giant  magnetostriction   was 
observed also in single crystals of other HTSC materials,\cite{Schmidt++} thus
establishing the effect as  being a general phenomenon which, like the  magnetization,
can provide important information about the pinning properties. 

Magnetostriction, as dilatations in general, are most frequently measured by displacive sensors like e.g. 
a capacitance dilatometer cell. Such sensors provide  an electrical signal which normally
represents the variation in the end-to-end dimensions of a sample.\cite{dilatometer}
Although very high sensitivities (10$^{-12}$ m) can be achieved, the cells are usually designed
 so that distortions in shape are not accounted for, or they may even influence the measurements
  in an uncontrolled way. Up to now,  shape distortions   in connection with 
the pinning-induced magnetostriction were never considered, and all data analysis has been based
on the  relation eq.(\ref{eq:1}) valid for plane strain deformations only. 

In this letter we show that the 
pinning-induced stress indeed leads to significant shape distortions 
even in the simple geometry of an infinitely long
isotropic cylinder with  square cross-section. 
For comparison we also present the exact solution of the magneto-elastic problem for the
circular cylinder where  shape is conserved.

From electrodynamics the force which drives the vorticies into the material is given
 by $ {\bf j} \times {\bf B} $ per unit volume.
For  HTSCs in the mixed state under  practical conditions
the current density, ${\bf j}$, is usually set equal to $({\bf \nabla} \times {\bf B})/\mu_0 $
as the thermodynamic field can be well approximated by ${\bf B}/\mu_0$.\cite{Zeldov-Clem}   In the
critical-state model one has $|{\bf j}| = j_c$, and  
a non-uniform distribution of pinned flux transmits onto the crystal
 a volume force density equal to $ {\bf f} = {\bf j}_c \times {\bf B} $.  
In the two geometries under consideration the symmetry of the force field is quite 
different, see Fig.1. In contrast to the simple radial distribution shown 
  in (a), the square cylinder in (b) is
 divided into four regions  each containing distributed  forces
 pointing the same direction. Moreover, in both cases 
the cylinder  generally has  additional divisions into shells where
the forces alternate in sign - or they can even vanish - all depending on the magnetic prehistory.
In all such states the  force exerted on the material per unit
volume is given by
\begin{equation}
   {\bf f} = ({\bf \nabla} \times {\bf B})\times {\bf B} / \mu_0
   =  - (2 \mu_0)^{-1} \;  {\bf \nabla} B^2 \; .
\label{eq:2}
\end{equation}
The last form,  expressing a purely magnetic free energy of the
 vortex lattice, is used in the present calculations.
In the analysis the strains are assumed well below the fracture limit 
allowing linear elasticity theory,
and hence the principle of superposition, to be applicable.

In the circular cylinder 
 the rotational symmetry implies that the deformation is described by
a radial displacement field, $u_r(r)$. The non-vanishing components of the
strain are $ e_r = u_r'(r) $  and $ e_{\theta} = u_r/r $,
which are related to the stress components,
$\sigma_{\theta}$ and $ \sigma_r$ by
$ E e_r = \sigma_r - \nu \sigma_{\theta} $ and
$   E e_{\theta} = \sigma_{\theta} - \nu \sigma_r $.
Here $E$ and $\nu$ is the Young's modulus and Poisson's ratio, respectively.

\centerline{\psfig{figure=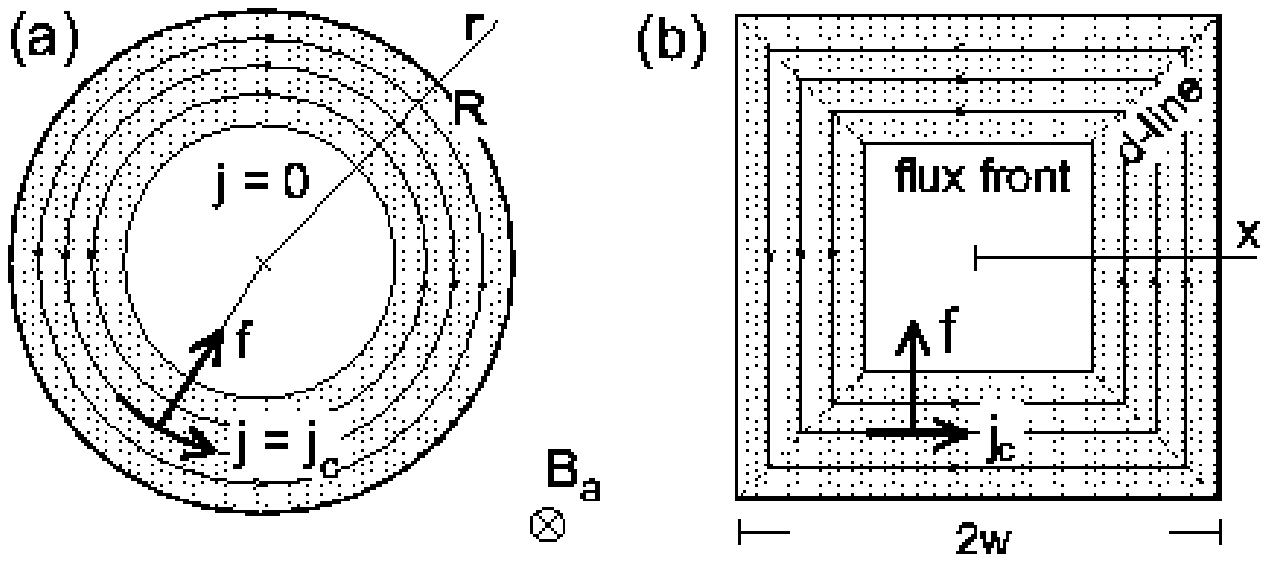,width=8.4cm}}

 {\small FIG. 1. Flux penetration and current flow pattern
  in a circular (a) and square (b) cylinder in the critical-state model. 
  The indicated magnetized states contain a flux front,  illustrating that
   the cylinders generally are divided  in shells with different volume forces.
 }\\[2mm]

The condition of static equilibrium is that 
$ (d \sigma_r/dr) + (\sigma_r -  \sigma_{\theta})/r + f =0 $,
which in terms of the displacement field is expressed as
\begin{equation}
   u_r'' + \frac{1}{r} \; u_r' - \frac{1}{r^2} \; u_r +
   \frac{1-\nu^2}{E} \; f(r) = 0 \ .
\label{eq:4}
\end{equation}
Integration twice, and using that $u_r(0) = 0$, gives
\begin{equation}
   u_r(r) = - \frac{1-\nu^2}{E} \; \frac{1}{r} \;
   \int_0^r  r' F(r') dr' +  C r \ ,                            
\label{eq:5}
\end{equation}
where  $  F(r) \equiv  \int_0^r f(r') dr' = [B^2(0) - B^2(r)]/2 \mu_0$.
With  the  constant $C$  determined by the free surface condition $\sigma_r(r=R) =0$,
the result becomes
\begin{eqnarray}
   u_r(r) & = & - \frac{1-\nu^2}{2 E \mu_0}  \; \left\{
   \left[ B_a^2 -  \frac{1-\nu}{R^2}  \int_0^R  r' B^2(r') dr' \right] \, r
   \right. \nonumber \\ && \left.
   - \ \frac{1+\nu}{r}  \int_0^r  r' B^2(r') dr' \right\} \ ,
\label{eq:6}
\end{eqnarray}
which is the  exact solution from which a complete stress-strain picture 
can be expressed in terms of the flux density $B(r)$. E.g., the radial stress is given by
\begin{eqnarray}
   \sigma_r(r) & = & \frac{1}{2 \mu_0} \; \left\{  B^2(r) - B_a^2 \ +
   \frac{1-\nu}{R^2} \left[ \int_0^R  r' B^2(r') dr'
   \right. \right. \nonumber \\     && \left. \left.
   - \ \frac{R^2}{r^2}  \int_0^r  r' B^2(r') dr' \right] \right\}  .
\label{eq:7}
\end{eqnarray}
The eqs.(\ref{eq:6}) and (\ref{eq:7}) generalizes  expressions for 
$u_r$ and $\sigma_r$ derived recently\cite{Ren} only for the Bean model, $j_c(B) = $ constant. 
We  emphasize that our results are  valid for any  critical-state model.

For the external dilatation, $\Delta R/R = u_r(R)/R $, one obtains  from eq.(\ref{eq:6}) 
\begin{equation}
    \frac{\Delta R}{R} =
    \frac{1- \nu}{E \mu_0  R^2} \ \int_0^R  r \; [B^2(r)-B^2_a] \; dr \ .
\label{eq:8}
\end{equation}
This formula for the pinning-induced magnetostriction of a circular cylinder,
first found by Johansen et al.,\cite{Riccione},
is  very similar to the expression in eq.(\ref{eq:1}) for the  slab. 
The extra factor $r$ in the integral is a simple modification, which one finds
analogously in the expressions for
 the critical-state magnetization. The factor  $1- \nu$ 
reflects the stronger constraint on free expansion for the cylinder as compared to the slab.


\centerline{\psfig{figure=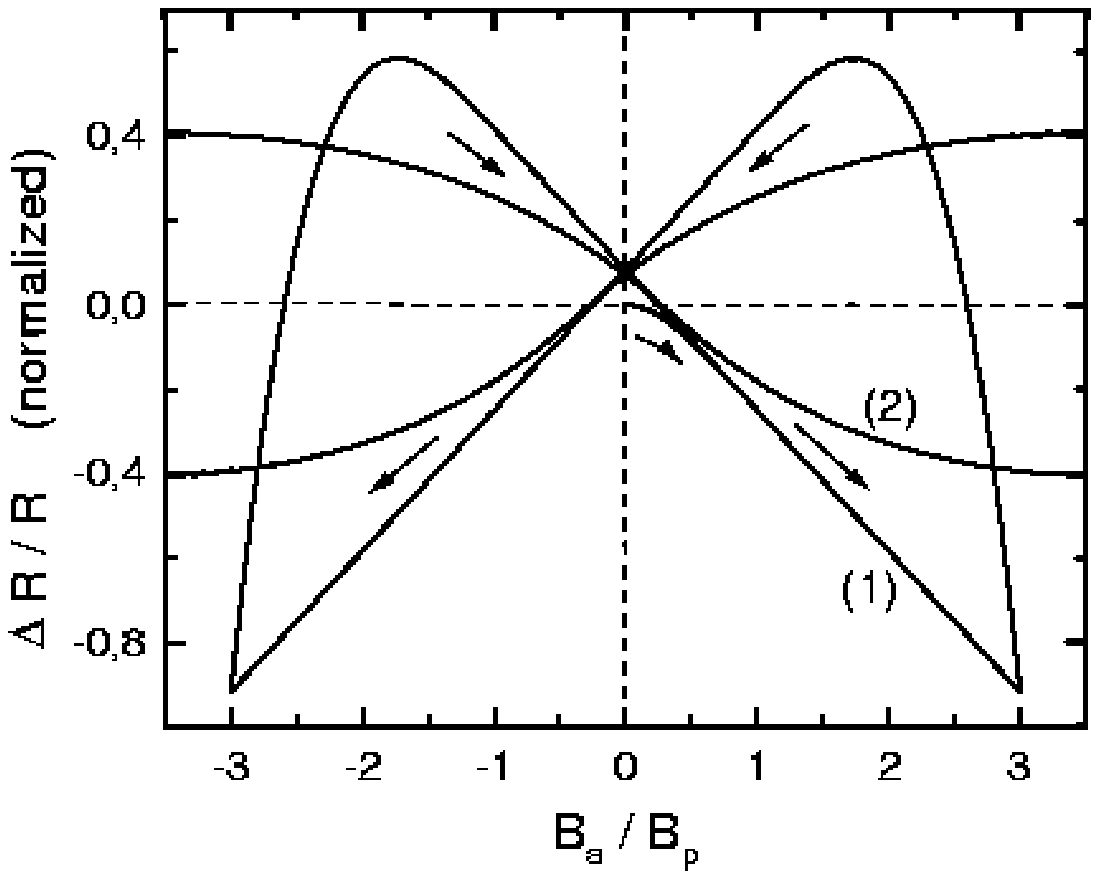,height=7.2cm}}

 {\small FIG. 2.
Magnetostriction, $\Delta R/R \times (1-\nu) (\mu_0 j_0 R)^2/E \mu_0$,
 as function of applied field  for a circular cylinder with $ j_c = j_0 \exp(-|B|/B_0) $. Curve (1);
$ B_0 = \infty$ i.e. the Bean model and (2); $ B_0 = \mu_0 j_0 R / 0.3$. The field axis is normalized
 by the full penetration field $ B_p = B_0 \ln(1+ \mu_0 j_0 R/B_0)$.
}\\[2mm]

From eq.(\ref{eq:8})  analytical expressions for the various
branches of magnetostriction hysteresis loops are readily derived. 
Figure 2 shows how $\Delta R/R$ depends on the applied field for two
commonly used critical-state models. 
As  $B_a$  runs through a complete cycle $\Delta R/R$ displays a huge
hysteresis loop  where in the Bean model case both the main ascending and descending branches
are linear provided a sufficiently high maximum field. 
Among other features of the Bean model results is that 
the maximum  remanent magnetostriction is given by
\begin{equation}
(\Delta R/R)_{\rm rem} =
     (1- \nu) (12 E)^{-1} \, \mu_0 \, j_c^2 \, R^2   \ .
\label{eq:9}
\end{equation}
Since the remanent state has no contribution from   reversible components,
the above relation provides a means to infere $j_c$ from magnetostriction
measurements. In the field range where the main branches
are linear the vertical width of the hysteresis loop  equals  
\begin{equation}
 (\Delta R/ R)_{\downarrow} -  (\Delta R/R)_{\uparrow}
 =  2 \, (1- \nu) (3 E )^{-1} \, R \, j_c \, B_a   \ .
\label{eq:10}
\end{equation}
 Note in particular  that if the hysteresis
width is normalized by $(\Delta R/R)_{\rm rem}$ the value
of $j_c$ can be found from the slope in a plot
 versus $B_a$ without knowledge of the elastic constants.
The normalized width is simply  $8 B_a/\mu_0 j_c R$, a result analogous
to the extensively used Bean model relation between $j_c$ and the width of
the magnetization loop.
Conversely will  measurements of magnetostriction 
 allow determination of the elastic parameter $(1-\nu)/E$
if  $j_c$ is known from other measurements.

The more curved  hysteresis loop resulting from the exponential model, 
 $j_c =j_0 \exp[-|B|/B_0]$, graph (2) in Fig.2, is usually a better description
of observed magnetostrictive behavior in HTSCs.\cite{Schmidt++,Ikuta-2}. An extensive
analysis with various  $j_c(B)$-functions is subject of a forthcoming paper,
where also the rather lengthy
expressions for the different branches in Fig.2, is presented.

A final remark on the  circular case is that the 
integral in eq.(\ref{eq:8}) also can be written as $\int_0^R   r^2 f(r) dr$, i.e., 
the $\Delta R/R$ is given as the second moment of the volume force distribution
just as the $j_c$-loops determine the magnetization. 
One could be tempted to generalize the relation  to other
cylindrical geometries, and thus allow for a simple unified treatment as done 
in magnetization calculations.\cite{THJ} However, such generalization would be incorrect, 
as will become evident as we now turn to the square cross-section case.

A different approach is chosen to
analyze the magneto-elastic problem of the square cylinder.
 Within the linear elasticity approximation one can
separate the treatment into an infinite number of  deformations produced by
infinitessimal square loops of the  force field (Fig.1 b). The final result is 
obtained by superimposing the elastic response of the whole cylinder  to each
force loop. For this to be a useful approach one needs
a way of adding these  deformations in a coherent manner. We first show that this
indeed is feasable for the square cylinder.

Consider the effect of a square force loop of width $dx$ with sides a distance $x$
from the origin (Fig.3). The material enclosed by the
loop will  experience only a  normal stress, and the deformation becomes 
a plane strain. The elastic response of the enclosed area
is  given by
$ E e_1 = \sigma_1 - \nu \sigma_2 $ and
$    E e_2 = \sigma_2 - \nu \sigma_1 $,
where $e_1$, $e_2$ and $\sigma_1$, $\sigma_2$ are the  plane strain
and normal stress components, respectively.
It immediately follows that the relative change in  the 
square area equals
$ e_1 + e_2  = - (1-\nu) E^{-1} \, 2 p $
where $p = - (\sigma_1 + \sigma_2)/2$ is the two-dimensional pressure.
The enclosed area  deforms under this pressure into another square so that a constant
gap $\delta g$ is created along the loop. Using that $p = f(x) dx$, the size of this  virtual
gap becomes 
\begin{equation}
     \delta g  =  (1-\nu)E ^{-1}  x f(x) dx \ .
\label{eq:11}
\end{equation}
The sign of $\delta g$ is so that a positive value represents a void space, whereas a negative
$\delta g$ corresponds to an expansion of the inner  square.

The contribution from this elementary deformation to the  strain
of the whole body is obtained by recombining\hfill   \newpage

 
\centerline{\psfig{figure=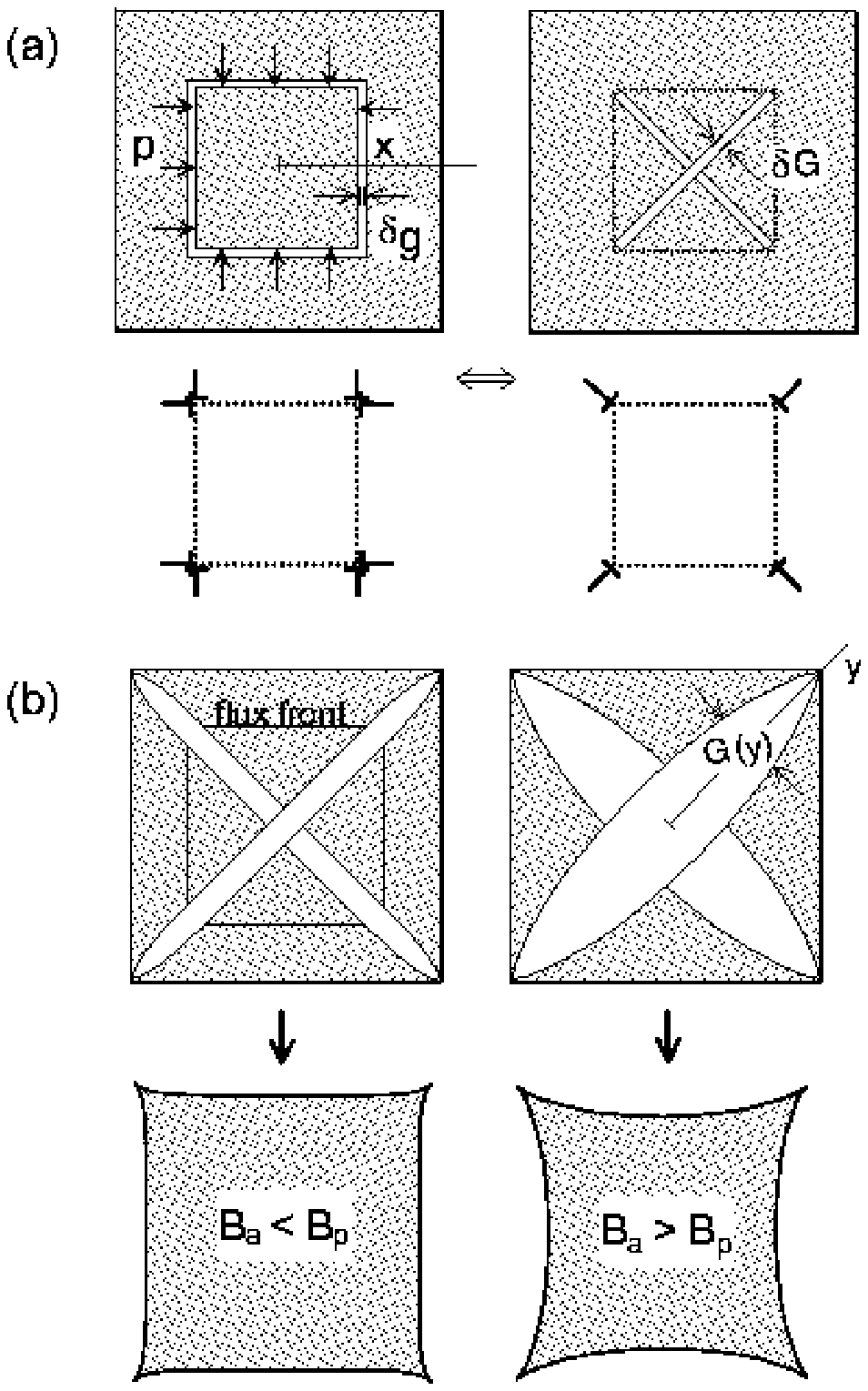,height=13.5cm}}

{\small FIG. 3.
(a) Partial effect of the volume force  in a square cylinder. By the shown 
transformation  the elastic problem can be treated analytically using 
 superposition, see text. (b) Schematic picture of the  deformation (strongly amplified) of
a square cylinder in the case of (left) incomplete and (right) complete flux penetration.
} \\[2mm]

\noindent
("gluing together") the parts on each side of the virtual gap. 
 The resulting strained state is equivalent to having a pair  of edge dislocations
in each  corner of the loop. As each pair 
can be replaced by one dislocation directed along the diagonal, we  conclude that
the original gap space can be replaced by  two slits of width
 $\delta G =  \delta g \sqrt{2}$ forming a central cross.
By this transformation  the integral effect of the entire  force
field can be determined analytically as follows. 
Let $y$ denote the distance from the 
center to a point on a diagonal. At this point the
total gap,  $G $, is an accumulation of the  gaps $\delta G$ created 
from $y$ and out to the corner $y= w \sqrt{2}$. The  total result is therefore given by
\begin{equation}
     \frac{G(y)}{w\sqrt{2}} =  \sqrt{2} \, \frac{1-\nu}{E \mu_0} \; \int_{y/\sqrt{2}}^w \, B(x) B'(x) x  \, dx  \ .
\label{eq:12}
\end{equation}
The full deformation is now obtained by "gluing" the virtually disrupted material together.
Since $G'(y) \propto   B j_c y$ it  follows that 
$G(y)$  is not  a linear function, thus proving that the pinning-induced magnetostriction
does not conserve shape in the square case.


\centerline{\psfig{figure=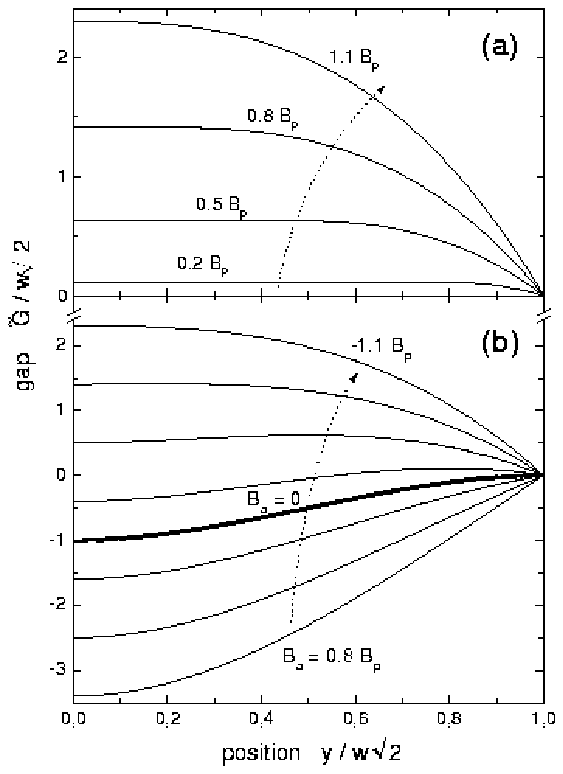,height=9.5cm}}

 {\small FIG. 4. Gap shape at various stages of flux penetration. (a) Increasing field
  ($B_a$ indicated) and (b) decreasing field for $B_a/B_p$ = 0.8, 
 0.5, 0.2, 0 (thick line), -0.2, -0.5, -0.8 -1.1.}\\[2mm]

The eq.(\ref{eq:12})  allows the gap to be  calculated.
In the Bean model along the increasing field branch one finds  for $\tilde{G} \equiv
G \, E \mu_0 (1-\nu)^{-1} B_p^{-2}$ the expression
\begin{equation}
 \frac{ \tilde{G}(y)}{w\sqrt{2}} = 
 3  [1- (\frac{y}{w \sqrt{2}})^2 ] \, (\frac{B_a}{B_p} -1) +
   2 - 2(\frac{y}{w\sqrt{2}})^3  ,
\label{eq:13}
\end{equation}
in the penetrated region. In the central Meissner area
the gap is constant.
The behavior is shown in Fig.4 a, and representative examples of the
resulting concave deformations are illustrated qualitatively in Fig.3 b. 
For the fully penetrated state one sees from  eq.(\ref{eq:13}) 
that $G(y)$ grows  linearly with $B_a$, while at the same time $G = 0$ at the corners.
This  produces  a deformation where the concave character grows
steadily with the applied field.

The behavior of $\tilde{G}(y)$ for descending applied field is shown in  Fig.4 b. 
The deformation along this branch  is evidently more diverse even when we omit the 
stage following immediately after  field reversal. The shown graphs represent only stages where
$j_c$ is reversed throughout the sample. The profiles with negative $\tilde{G}$ gives rise to 
convex deformations. The convexity  persists all the way down to the remanent state,
where the distortion near the corners is vanished. Then, as the field is reversed there is an
 interval where the behavior is governed by a $G(y)$ with an oscillatory shape, thus giving a
mixed type of deformation. As the reverse field increases in magnitude the shape again becomes concave.
For $B_a \leq -B_p$ the sample shape follows the same development as during the field increase. 

In conclusion, we have investigated and compared the pinning-induced magnetostriction
in two important geometries. In contrast to the shape-conserving circular case,
which could be solved exactly, the existence of current discontinuity(d)-lines
 in the square case  complicates the behavior by producing shape distortion. 
We have predicted the appearance of  convex, concave and even more complicated types of deformations
where the distortion is of the same order of magnitude as the overall striction. 
Since in most cases HTSC crystals  have a rectangular shape, which always lead to 
d-lines in the critical-state, the  distortion could easily lead to misinterpretation  of
dilatometric measurements. The problem should be minimized by (i) having a high aspect ratio
rectangular shape  so that the central part of the sample is described approximately by a plane
 strain deformation, and (ii)  designing the displacive sensor with a local contact to the
long side of the rectangle in order to avoid distorted corner regions. 
Full analysis of the proper corrections will clearly require numerical work.


The authors acknowledge  discussions with Yu. Galperin, and financial
  support from  The Research Council of Norway.\\

\end{document}